\documentclass[letterpaper,9pt, twocolumn]{extarticle}
\usepackage{fixltx2e}[2006/03/24]  
\usepackage[letterpaper, dvips=true, scale=0.86, nofoot, centering ]{geometry}
\usepackage[T1]{fontenc}
\usepackage[note,mcite]{achemso}
\usepackage{amssymb}
\usepackage{amsmath}
\usepackage[dvips]{graphicx}

\newcommand{\chemeq}[2]{\underset{#2}{\stackrel{#1}{\rightleftharpoons}}}

\makeatletter
\renewcommand\section{\@startsection {section}{1}{\z@}%
       {-1.5ex \@plus -.5ex \@minus -.8ex}%
       {1.5ex \@plus.2ex \@minus .2ex}%
       {\raggedright\normalfont\large\bfseries\sffamily}}
    \renewcommand\subsection{\@startsection{subsection}{2}{\z@}%
       {-1ex\@plus -.4ex \@minus -.4ex}%
       {1ex \@plus .2ex \@minus .2ex}%
       {\raggedright\normalfont\bfseries\sffamily}}
    \renewcommand\subsubsection{\@startsection{subsubsection}{3}{\z@}%
      {2ex \@plus1ex \@minus.3ex}%
      {-1em}%
      {\normalfont\normalsize\bfseries\sffamily}}
\makeatother
\title{Microreversible recycled chemical systems. Comment on ``A
  Re-Examination of Reversibility in Reaction Models for the
  Spontaneous Emergence of Homochirality''}

\author{\emph{Raphaël Plasson}\thanks{\texttt{rplasson@nordita.org}}\\
           Nordita, Stockholm, Sweden}

\date{}

\begin{document}

\maketitle
\thispagestyle{empty}


The question of the onset of the homochirality on prebiotic Earth
still remains a fundamental question in the quest for the origin of
life\cite{palyi-02,*palyi-04}. Recent works in this field introduce
the concept of recycling\cite{saito-04,plasson-04,plasson-07}, rather
than the traditional open-flow system described by
Frank\cite{frank-53}. This approach has been criticized by Blackmond
\emph{et al.}\cite{blackmond-08}. They claimed that such systems are
thermodynamically impossible, except in the cases where
non-microreversible reactions are introduced, like in photochemical
reactions, or under the influence of physical actions (e.g. by crystal
crushing\cite{viedma-05}). This point of view reveals
misunderstandings about this model of a recycled system, overlooks the
possibility of energy exchanges that could take place in prebiotic
systems, and leads the authors to unawarely remove the activation
reaction and energy source from their ``non-equilibrium'' models. It
is especially important to understand what are the concepts behind the
notion of recycled systems, and of activation reactions. These points
are fundamental to comprehending how chemical systems --- and
especially prebiotic chemical systems --- can be maintained in
non-equilibrium steady states, and how free energy can be used and
exchanged between systems. The proposed approach aims at the
decomposition of the problem, avoiding to embrace the whole system at
the same time.

\begin{figure}[h]
  \centering
  \includegraphics[width=8cm]{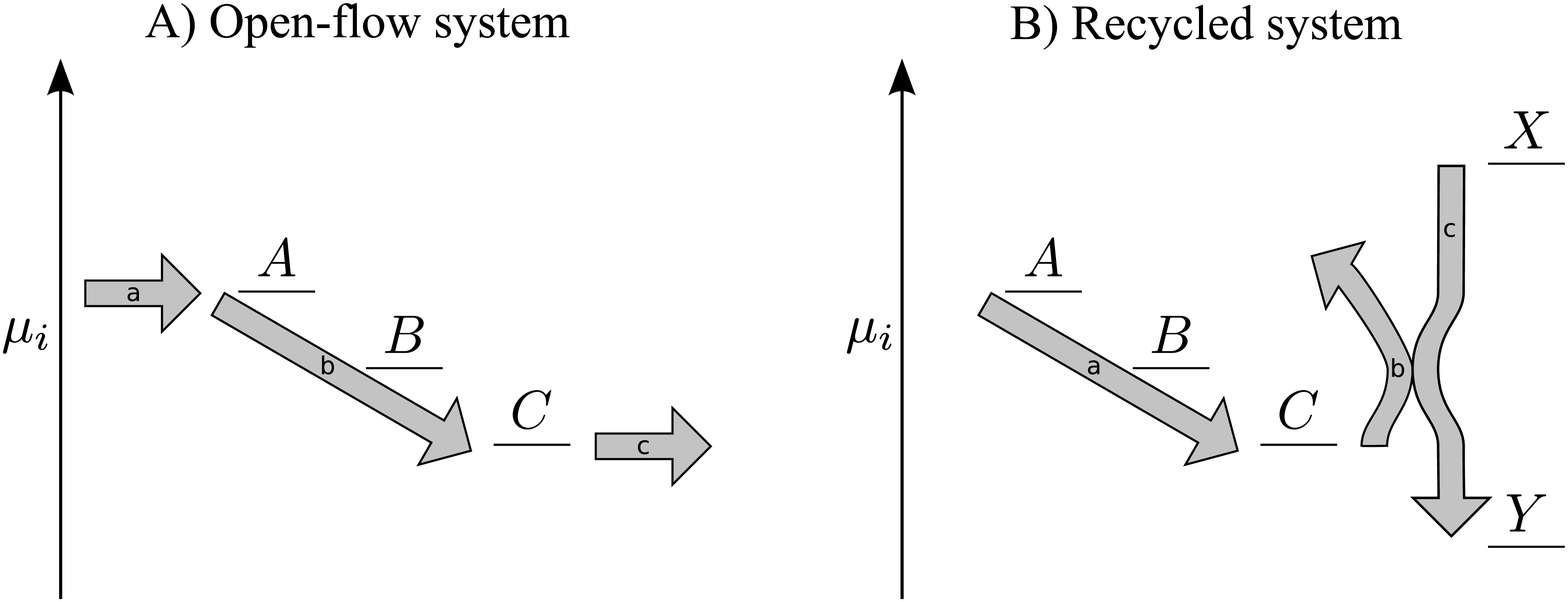}
  \caption{Schematic representation of open-flow and recycled
    systems. A: a, input flow of compound $A$; b, spontaneous
    transformation from $A$ to $C$ through $B$; c, output flow of
    compound $C$. B: spontaneous transformation from $A$ to $C$
    through $B$; b, activated transformation of $C$ into $A$; c,
    spontaneous transformation of $X$ into $Y$, coupled to the
    transformation b.}
  \label{fig:activ}
\end{figure}

In an open flow system, the difference in chemical potential of the
studied compounds is maintained by continuous input and output flows
of these compounds (Fig.~\ref{fig:activ}.A). In a recycled system,
some low potential compounds are activated back into high potential
compounds by an external process (Fig.~\ref{fig:activ}.B). If this
recycling results in maintaining the studied compounds in a closed
system, it must also be coupled to the consumption of free energy,
i.e. the continuous consumption of ``fuel'' external compounds and
disposal of ``waste'' compounds. Recycled systems are not --- and have
never claimed to be --- strictly closed. In the case of the
Activation\slash{}Polymerization\slash{}Epimerization\slash{}Depolymerization
system (APED)\cite{plasson-04,plasson-07}, the theoretical model can
be mathematically reduced to a closed system of amino acid residues,
by impliciting the sources of energy in the kinetic parameters. But
when referring to the chemical counterpart of the theoretical
reactions, it is clearly stated that this system is linked to the
consumption of additional reactants\cite{rem-plasson}.

The process of recycling aims at maintaining a difference in chemical
potentials inside the system. In the case described in
Fig.~\ref{fig:activ}.B, the reaction is spontaneous from $A$ to $B$
and from $B$ to $C$, and thus from $A$ to $C$. As a consequence,
forcing the reaction from low potential compounds ($C$) to high
potential compounds ($A$) requires an active process, linked to the
consumption of energy. In the case of chemical energy, this means that
external compounds of high chemical potential ($X$) are consumed, and
low chemical potential compounds ($Y$) are being disposed of. As no
$A$, $B$ or $C$ are never entering nor leaving the system, these three
compounds being only interconverted from one to another, the
$\{A,B,C\}$ system can be considered as closed, and the sum of the
concentrations $[A]+[B]+[C]$ is strictly constant. However, it is
coupled by a reaction $A+X \leftrightharpoons B+Y$ to the external
system $\{X,Y\}$, that is opened so that the difference in
chemical potentials is maintained between these two compounds.  An
activation reaction does not consist in a simple spontaneous reaction,
but rather represents a non-spontaneous reaction that is made feasible
thanks to the coupling with another reaction, that is spontaneous with
a larger difference in chemical potential. It is thus possible to
obtain a transfer of free energy, that forces the activated reaction
from stable compounds to unstable ones, as in many endergonic
biochemical reactions coupled for example to ATP hydrolysis.

\begin{figure}[h]
  \centering
  \includegraphics[width=8cm]{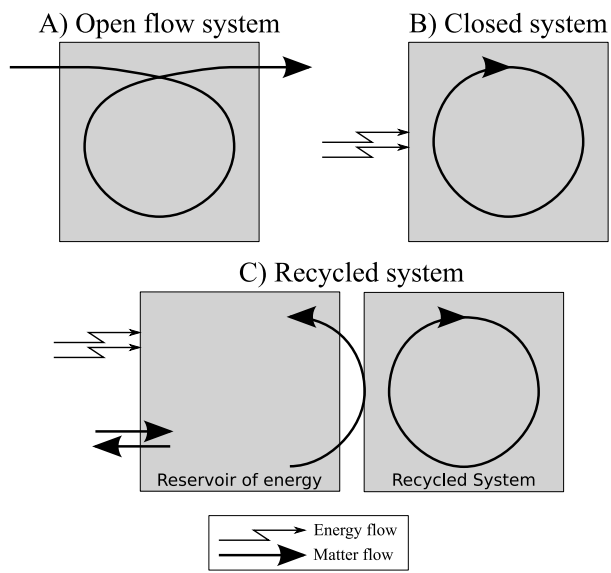}
  \caption{Representation of the decomposition of non-equilibrium
    systems into subsystems, emphasizing the different exchanges
    between subsystems and the surrounding.}
  \label{fig:recycle}
\end{figure}

In that scope, the assumption that the system is recycled comes from
decomposing the whole system into several interconnected
subsystems, rather than just stating that the system is globally open,
i.e. that some matter is exchanged with the surroundings
(Fig.~\ref{fig:recycle}.A), or globally closed, i.e. that the system
only exchanges energy but not matter (Fig.~\ref{fig:recycle}.B). The
recycled system is described as closed for some given internal
compounds (internal system), but linked to the consumption of
compounds from a ``reservoir'' (external system), matter flow being
established as a link between these two systems. The mechanisms that
maintains the state of the reservoir are implicit, relying on other
processes that are not related to the internal processes of the
recycled system (Fig.~\ref{fig:recycle}.C). The environment is
described as a black-box, that is maintained in a non-equilibrium
state by implicit external factors. The recycled system is just
``plugged'' to this environment acting as a reservoir of free energy.
The reservoir system does not need the recycled system to exist, as it
is maintained anyway by external sources. But the recycled system
relies on the presence of the reservoir, to be maintained in an active
state. The reservoir preexists, as an initial requirement to the
possibility for a more complex non-equilibrium system to be
established.

The recycled system is macroscopically irreversible, even if composed
of microreversible reactions. The origin of this irreversibility is
simply the transfer of free energy, and namely chemical free energy,
generating unidirectional cycles of reaction. For example, a reaction
$A\chemeq{k_1}{k_{-1}} B$ can be activated, linked to the
transformation of $X$ and $Y$ compounds through the reaction $A+X
\chemeq{k_a}{k_{-a}} B+Y$.  
The activated reaction is mathematically
equivalent to a $A\chemeq{k'_1}{k'_{-1}} B$ reaction with apparent
kinetic rates $k'_1=k_ax$ and $k'_{-1}=k_{-a}y$, $x$ and $y$ being
maintained in non-equilibrium concentration in the reservoir
system. The sole intrinsic kinetic constants follow the thermodynamic
relationship ($\frac{k_1}{k_{-1}}=K_1$), while the apparent kinetic
constant do not ($\frac{k'_1}{k'_{-1}}=K_a\frac{x}{y}$), as they are
function of the concentrations of $X$ and $Y$. The condition
$\frac{k'_1}{k'_{-1}}=K_1$ will be only obtained in the case where $X$
and $Y$ would be at equilibrium concentrations. It is thus possible to
write reactions whose kinetic parameters do not correspond to
thermodynamic parameters, as long as it is kept in mind that this is
hiding a source of energy and the possible implicit consumption of
external compounds. The reaction can of course also be directly
activated in one specific direction by a photochemical
reaction\cite{photochem}.


In the case of the APED system\cite{plasson-04}, the activation
reaction is based on the transformation of non-activated amino acid
$L$ (or $D$) into an activated form noted $L^*$ (or $D^*$) by a
reaction $L\rightarrow L^*$. As previously stated, this theoretical
activation reaction can represent the transformation of an amino acid
into its N-carboxyanhydride form (NCA), which is an activated reaction
that relies on the consumption of reactive
compounds\cite{kricheldorf-87,taillades-99,huber-98,plasson-07}, and
is in no case a ``thermal activation'' as stated by Blackmond \emph{et
  al}\cite{blackmond-08}. The global formula is $L+X
\underset{k_{-a}}{\stackrel{k_a}{\rightleftharpoons}} L^*+Y$. The
compound $X$ corresponds to a high potential compound, and $Y$ to a
low potential compounds\cite{rem-nca}, so that the reaction constant
$K_a$ is somewhat larger than one. Moreover, the molecules $X$ and $Y$
are maintained constant, so that the reaction comes down to the first
order reaction $L\leftrightharpoons L^*$ of apparent kinetic rates
$a=k_a[X]$ and $a_-=k_{-a}[Y]$. The relationship with the
thermodynamic constant is $\frac{a}{a_-}=K_a\frac{[X]}{[Y]}$. In order
to guarantee $a\gg a_-$, it is thus sufficient to guarantee 
both $K_a
\gg 1$ (i.e. $X$ is a reactive compound while $Y$ is a stable
compound) and $[X] \gg [Y]$ (i.e. the system is continuously fed with
$X$ while $Y$ is dissipated).  The spontaneous deactivation is also
taken into account, but does not correspond to the microreversible
inverse of the activation reaction.  This process is the simple
hydrolysis of the NCA back to the amino acid, releasing carbon
dioxide: $L^*+ \mbox{H}_2\mbox{O}
\underset{k_{-b}}{\stackrel{k_b}{\rightleftharpoons}} L +\mbox{CO}_2$.
The apparent kinetic rate of the reaction $L^*\leftrightharpoons L$
are $b=k_b$ and $b_-=k_{-b}[\mbox{CO}_2]$. In accordance with the well
known reactivity of NCAs\cite{kricheldorf-87}, the reaction of
deactivation is totally displaced toward the formation of amino acid,
i.e. $k_b \gg k_{-b}$, so that $b \gg b_-$.

The model previously described\cite{plasson-04,plasson-07} is thus a
correct approximation, as both $a \gg a_-$ and $b \gg b_-$. The ratio
between $a$ and $b$ is also not fixed, depending on $[X]$, here again
high concentrations of $X$ favoring the activation reaction. Moreover,
the spontaneous hydrolysis of NCA is slow relatively to other
reactions in appropriate conditions\cite{stab-nca,*plasson-02}.  In
the model described by Blackmond \emph{et al.}\cite{blackmond-08}, the
microreversible reaction of \emph{both} activation and deactivation
reactions should have been added, as explained above, but they
artificially considering that the activation is the microreversible
reverse reaction of the deactivation reaction. They should also have
taken into account the fact that energetic compounds are consumed
during the activation reaction, but they confused intrinsic kinetic
rates with apparent kinetic rates, leading to remove the activating
compounds from the system. They thus actually \emph{removed} the
activation reaction. There is thus no surprise that the system fails
to evolve towards a non-racemic steady state in absence of source of
energy. The system does not in any way describe the full APED model.


Maintaining a non-equilibrium chemical system in a non-racemic
steady-state relies on maintaining unidirectional cycles of reaction
by consuming free energy. The description of such chemical cycles were
introduced very early in the origin of life theory by
G\'anti\cite{ganti-75,*ganti-97} and Eigen\cite{eigen-71,*eigen-77}.
They are fundamental for the self-organization of matter, as they can
induce network autocatalytic properties. The consumption of energy
implies that reactions are not detailed balanced: they are proceeding
faster in one direction than in the other, even taking into account
the microreversibility, driven by the differences in chemical
potentials. It is important to distinguish the mechanism of the
deracemization process from the energetic source allowing this process
to operate. While this distinction is not done in the traditional
approach\cite{frank-53}, it appears that the open flow system is the
source of energy, and the recycled system is responsible for the
network autocatalysis. The purpose of the previous
articles\cite{plasson-04,plasson-07} was to focus on the autocatalytic
mechanism itself, and to understand how it works. The fact that energy
can flow inside the system was already a known fact, addressed in the
study of the potential activation of amino acid into NCAs in prebiotic
conditions\cite{taillades-99,huber-98}, and was thus let as implicit
in the equations (but not in the discussions). The last point is now
to understand how the free energy is distributed and used inside the
system. This point will be addressed in details in a future
publication.

\small

\paragraph{Acknowledgement:} I thank H.~Bersini, R.~Pascal and
A.~Brandenburg for discussions and comments.


\begin{mcitethebibliography}{10}
\bibitem{photochem}
A photochemical reaction is of first order in the reactant. The exponential
  term of the Beer-Lambert law is only accounting for the decrease of the
  intensity of the incident light while it is penetrating inside the reactor,
  thus the exponential term proportional to the reactor width $\ell$. For
  systems of low width, for example in the case of a photochemical reaction
  taking place in a shallow pond with an irradiation coming from the top, the
  radiation intensity is uniform inside the system, and the Beer-Lambert law
  reduce to a first order reaction, cancelling the term $\ell$.\relax
\mciteBstWouldAddEndPunctfalse
\mciteSetBstMidEndSepPunct{\mcitedefaultmidpunct}
{}{\mcitedefaultseppunct}\relax
\EndOfBibitem
\bibitem{rem-nca}
For example we can have $X=\mbox{CO}$ and $Y=\mbox{H}_2$\cite{huber-98}, or $X
  = \mbox{HNCO} + \mbox{N}_2\mbox{O}_3$ and $Y = \mbox{N}_2 + \mbox{HNO}_2 +
  \mbox{H}_2\mbox{O}$\cite{taillades-99}.\relax
\mciteBstWouldAddEndPunctfalse
\mciteSetBstMidEndSepPunct{\mcitedefaultmidpunct}
{}{\mcitedefaultseppunct}\relax
\EndOfBibitem
\bibitem{stab-nca}
For example, it could be determined experimentally that the NCA of valine is
  stable enough towards hydrolysis to survive several hours in water at $pH~6$
  and $5~^o$C, its coupling with amino acids being much faster\relax
\mciteBstWouldAddEndPuncttrue
\mciteSetBstMidEndSepPunct{\mcitedefaultmidpunct}
{\mcitedefaultendpunct}{\mcitedefaultseppunct}\relax
\EndOfBibitem
\end{mcitethebibliography}
\onecolumn

\ifx\mcitethebibliography\mciteundefinedmacro
\PackageError{achemsoM.bst}{mciteplus.sty has not been loaded}
{This bibstyle requires the use of the mciteplus package.}\fi

\end{document}